\RequirePackage{fix-cm}
\documentclass[smallextended]{svjour3}
\smartqed
\usepackage{graphicx}
\usepackage{subfigure}
\usepackage[misc]{ifsym}
\usepackage{xcolor}
\usepackage{capt-of}

\usepackage{amsmath}
\usepackage{amssymb}
\usepackage{subfigure}
\usepackage{multirow}
\usepackage{epstopdf}
\usepackage{color}

\begin{document}

\title{Multivariate Spatial Data Visualization: A Survey}
\subtitle{}
\author{Xiangyang He\and
        Yubo Tao\and
        Qirui Wang\and Hai Lin
}

\institute{Xiangyang He\and Yubo Tao(\Letter) \and Qirui Wang \and Hai Lin \at
            State Key Laboratory of CAD$\&$CG, Zhejiang University, Hangzhou, China\\
            \email{xiangyanghe@zju.edu.cn, taoyubo@cad.zju.edu.cn, qiruiw@gmail.com, lin@cad.zju.edu.cn}
}

\date{Received: 30 July 2018 / Accepted: 18 August 2018}
\maketitle

\begin{abstract}
Multivariate spatial data plays an important role in computational science and engineering simulations. The potential features and hidden relationships in multivariate data can assist scientists to gain an in-depth understanding of a scientific process, verify a hypothesis and further discover a new physical or chemical law. In this paper, we present a comprehensive survey of the state-of-the-art techniques for multivariate spatial data visualization. We first introduce the basic concept and characteristics of multivariate spatial data, and describe three main tasks in multivariate data visualization: feature classification, fusion visualization, and correlation analysis. Finally, we prospect potential research topics for multivariate data visualization according to the current research.

\end{abstract}

\section{Introduction}
\label{intro}

Multi-physics simulations are trending in the modern scientific computation, and these simulations often generate data sets with multiple variables related to complex physical phenomena in physical space. These data sets are generally called multivariate data or multifield data. For each spatial point $p$ in physical space, its associated values are $\{v_1(p), v_2(p), ..., v_n(p)\}$, where each variable $v_i$ represents a physical or chemical property, and $n$ is the number of variables of the multivariate data.
Each variable of multivariate data could be a scalar, vector , or tensor field. Because it has multiple variables, multivariate data is heterogenous and complex, making it difficult to visualize, analyze, and gain insights from the data.
Multivariate data can be used to extract hidden relationships and explore existing phenomenona or new laws in many applications, such as computational fluid dynamics, electromagnetic field simulation, combustion simulation, and meteorological simulation. For example, three-dimensional (3D) meteorological data can be obtained by computer simulations including multiple variables, such as temperature, pressure, and humidity. Users can intuitively explore the correlations or features between these variables and even discover new phenomena and laws. Therefore, multivariate spatial data visualization has always been an important research topic in the field of scientific visualization.

With the improvement in the computational performance of supercomputers, the simulation range can be very large, and the achievable resolution is also very high. Multifield visualization is one of the top scientific visualization research problems~\cite{Johnson:2004:TSV}, owing to its complex structures and inner relationships. The goal is to effectively visualize multiple variables simultaneously and intuitively represent their mutual interactions.

\begin{figure}[t]
\centering
\includegraphics[width=0.6\columnwidth]{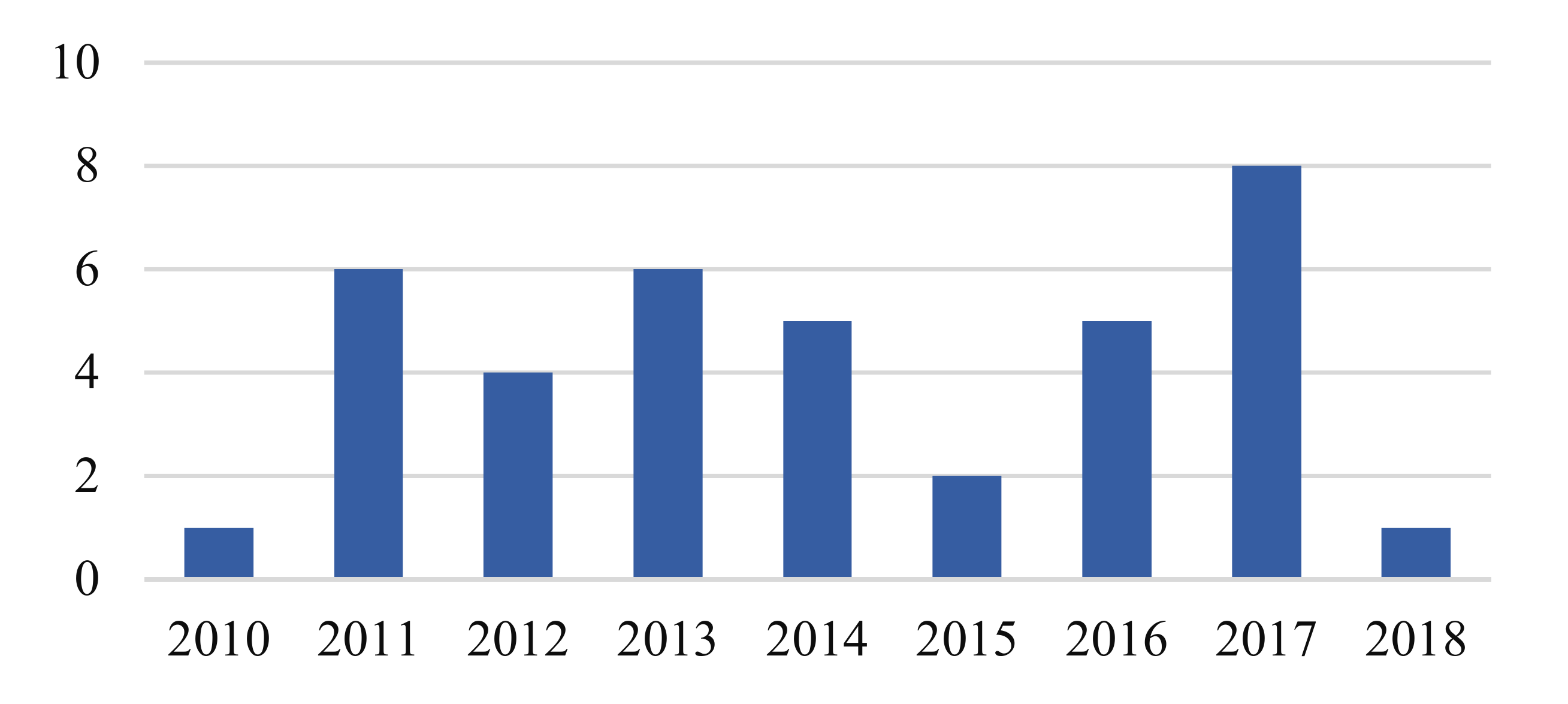}
\caption{Number of publications related to multivariate data since 2010.}
\label{fig:sec1_0}
\end{figure}

The main research problems of multivariate data visualization are feature classification, fusion visualization, and correlation analysis. In feature classification, features can be classified based on a scalar value and its derived attributes for univariate data, whereas they are usually defined by multiple variables for multivariate data. It is important and necessary to use multiple variables to define and classify features for complex phenomena. In fusion visualization, variables can be rendered separately in juxtaposed views, but they fail to intuitively establish spatial relationships between multiple variables. Thus, it is necessary to visualize multiple variables simultaneously in the same physical space to facilitate the comparison of multiple features and the understanding of their interaction among multiple variables. In correlation analysis, multivariate data generally can have hidden associations, because they work collectively in the simulation model~\cite{Carr:2014:JCN}. For instance, a hurricane is a rapidly rotating storm system characterized by a low-pressure center, strong winds, along with heavy rain in climate simulations. Thus, it would be beneficial to extract and represent their relationships via combining information visualization and scientific visualization methods from multiple aspects, such as the interactions among variables, numerical values, and features.

For the above three main research problems, we conduct a thorough literature review based on relevant works from major visualization venues, namely IEEE VIS, EuroVis, PacificVis, and the visualization journals, such as IEEE Transactions on Visualization and Computer Graphics (TVCG) and Computer Graphics Forum (CGF). Kehrer and Hauser~\cite{Kehrer:2013:VAV} summarized multiple facets of scientific data, including multi-dimensional, multivariate, multi-value, multi-modal, and multi-model aspects published prior to 2012. In this paper, we mainly focus on multivariate data visualization, especially the recent progress in feature classification, fusion visualization, and correlation analysis. Fig.~\ref{fig:sec1_0} shows the number of publications related to multivariate data since 2010. The histogram is not claimed to be complete, but reflects a trend showing that multivariate data visualization has been a hot topic in recent years. Unlike other similar reviews, this paper aims to fill the gap in the literature review in the field of multivariate data visualization since 2010 and provides an up-to-date and detailed overview of the recent advances. In this paper, we first introduce feature classification methods for multivariate data. Next, we discuss the fusion visualization based on different stages of the visualization pipeline. We then review the correlation analysis of variables, voxels, numerical values, and features. Finally, we provide several future research directions in multivariate data visualization. Table~\ref{tab:refernces} shows the categories of each of these sections and their corresponding references published after 2010.

\begin{table*}[bp]
    \centering
     \caption{Related references in this paper for each section published after 2010.}
     \label{tab:refernces}
    \begin{tabular}{c|c|c|c}
	 \hline
 	 & Sections & Categories & References \\
 	 \hline
\multirow{11}{2.5cm}{Multivariate spatial data visualization}
&
\multirow{3}*{Feature classification}
	 &  Interactive classification  &  ~\cite{Zhao:2010:MRA}~\cite{Zhou:2012:TFC},~\cite{Guo:2012:SMV},~\cite{Guo:2014:SLA},~\cite{Ljung:2016:SOT},~\cite{Liu:2016:AAF},~\cite{Zhou:2017:IPC},~\cite{Lu:2017:MVD}  \\ \cline{3-4}
	& &  Data mining  &   ~\cite{Hong:2014:FLDA},~\cite{Zhou:2014:GSS},~\cite{Soundararajan:2015:LPT},~\cite{Wu:2015:EAF},~\cite{Wang:2016:MPM},~\cite{Wei:2017:EDF}   \\ \cline{3-4}
	& &  Topological structures &  ~\cite{Nagaraj:2011:RIE},~\cite{Duke:2012:VNS},~\cite{Carr:2014:JCN},~\cite{Geng:2014:VAO},~\cite{Carr:2015:FSG},~\cite{Wu:2017:DMV},~\cite{Tierny:2017:JFS}   \\ \cline{2-4}
	
&
\multirow{3}*{Fusion visualization}
     &  Data fusion  &~\cite{Suthambhara:2011:SOJ},~\cite{Guo:2012:SMV},~\cite{Khlebnikov:2013:NVR},~\cite{Huettenberger:2014:DAS},~\cite{Carr:2015:FSG},~\cite{He:2017:RLT},~\cite{Lu:2017:MVD},~\cite{Dutta:2017:PIG}   \\ \cline{3-4}
	& &  Feature fusion &   ~\cite{Kuhne:2012:ADA},~\cite{Guo:2014:SLA},~\cite{Ding:2016:VIO}   \\ \cline{3-4}
	& &  Image fusion  &   ~\cite{Borgo:2013:GVF},~\cite{Zhou:2014:GSS},~\cite{Schroeder:2016:VAA}  \\ \cline{2-4}	

&
\multirow{5}*{Correlation analysis}
     &  Voxels &   ~\cite{Nagaraj:2011:AGC},~\cite{Wu:2015:EAF}   \\ \cline{3-4}
	& &  Variables  &   ~\cite{Wang:2011:AIT},~\cite{Biswas:2013:AIF},~\cite{Dutta:2017:PIG}   \\ \cline{3-4}
	& &  Numerical values  &   ~\cite{Haidacher:2011:VAU},~\cite{Huettenberger:2013:TMS},~\cite{Carr:2014:JCN},~\cite{Liu:2016:AAF}   \\ \cline{3-4}	
	& &  Features  &   ~\cite{Schneider:2013:ICO},~\cite{Wang:2017:FAV}   \\ \cline{3-4}
	& &  Value-variable  &   ~\cite{Biswas:2013:AIF},~\cite{Liu:2016:AAF}   \\ \cline{1-4}
\end{tabular}
\end{table*}

\section{Feature Classification}

With complex, large-scale, multivariate data, it is essential to locate important features in the data and analyze these features to better understand the relevant phenomena.
For example, a hurricane eye shows low pressure and medium wind speed, and the non-combustion region of the turbulent combustion data set shows a high heat release rate and low hydroxyl radical concentration. Features in multivariate data are relatively complex, and the relationships between variables are also hidden. Extracting and classifying these features is the first step in multivariate data visualization. Transfer functions are commonly used in scalar fields, and they can be extended to high-dimensional transfer functions for interactive classification. However, it is a time-consuming and challenging task to interactively classify features, because users need to perform ``trial-and-error'' work according to the statistical characteristics of the data. In this case, data mining can be employed to implement automatic or semi-automatic feature classification. Moreover, abstract topological representation methods such as isosurfaces and contour trees can be used to extract isosurfaces and segment features in scalar fields. They can also be used to extract and express features in multivariate data. Feature classification methods can extract and classify multiple features from raw data, such as the outer frame and the frame layer of the turbulent combustion data set. In this section, we classify the feature classification methods into three categories: interactive classification based on multiple variables, feature classification based on data mining and feature classification based on topological structures, and summarize the corresponding research.

\begin{figure}[hbp]
\centering
\includegraphics[width=0.65\columnwidth]{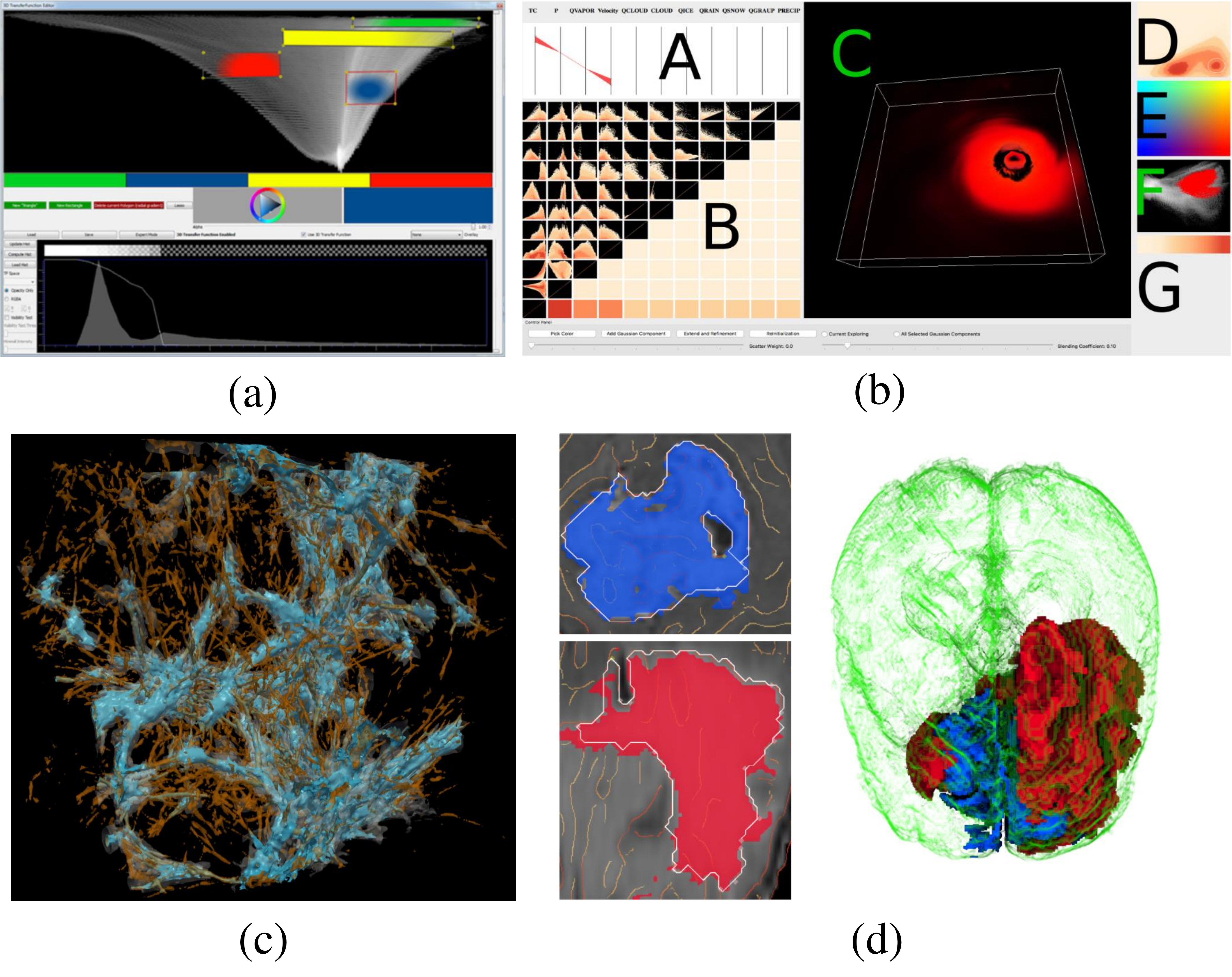}
\caption{Examples of feature classification. (a) Four features can be classified from a two-dimensional transfer function composed of pressure and temperature based on interactive feature classification~\cite{Zhou:2012:TFC}. (b) Interactive feature classification using a high-dimensional transfer function based on a scatter plot matrix~\cite{Lu:2017:MVD}. (c) Feature classification based on data mining~\cite{Zhou:2014:GSS}. (d) Isosurfaces of two variables and their fiber surfaces based on topological structures~\cite{Carr:2015:FSG}.}
\label{fig:sec2_1new}
\end{figure}

\subsection{Interactive Classification based on Multiple Variables}

Interactive classification based on multiple variables can statistically analyze multivariate data and present statistical information through information visualization. It can implement feature classification by selecting different attributes or variables interactively. In general, multivariate data can be transformed into a statistical space, such as a histogram-based transfer function,  parallel coordinate plot (PCP), and scatter plot (SP). It should be noted that transfer functions consider only partial variables, whereas PCP and SP can present all the variables.

Given a domain $x$ representing a set of points, each of which presents an n-dimensional vector containing scalar values of all variables with the same space coordinate, the mathematical expression of a transfer function can be defined as: $T: x\rightarrow\{c, \alpha, ...\}$, where $c$ is color and $\alpha$ is opacity. In addition, there are other ranges such as illumination coefficient and texture. Many different geometric attributes and statistical properties were proposed according to the characteristics of the extracted features. In 2016, Ljung et al.~\cite{Ljung:2016:SOT} systematically provided an up-to-date overview of transfer functions for the volume rendering of univariate data.
In general, a high-dimensional transfer function is defined for selecting features interactively, and then different colors are mapped to the corresponding features to show each distribution. For univariate data, high-dimensional transfer functions are derived by combining multiple attributes. However, it is more difficult to design a high-dimensional transfer function for multivariate data, owing to the complexity of the data, which requires more research and exploration. Gradients have been widely used in the design process of transfer functions, and they are generally composed of two-dimensional (2D) histograms with scalar values~\cite{Kniss:2002:MTF}.
Zhou and Hansen~\cite{Zhou:2012:TFC} supported discretionary combinations of one-dimensional (1D) and 2D transfer functions for realizing the hierarchical division of features and selecting multiple attributes to reduce the correlations between the dimensions of high-dimensional transfer functions, as shown in Fig.~\ref{fig:sec2_1new}(a).

Because there are usually more than ten variables in multivariate data, it is necessary to embed all the data onto a 2D visual plane by projection reduction methods to help users explore internal structures and distributions intuitively. A PCP is composed of multiple parallel axes, and each axis represents a variable of the multivariate data. Each spatial sampling point is presented on a PCP to show a polyline with vertices on the axes, and the position of the vertex on the $i$-$th$ axis represents the scalar value of the $i$-$th$ variable for this spatial point. PCPs are intuitive with respect to the distribution of each variable/dimension of multivariate data, owing to the presentation of numerical information of all variables. Users can define numerical interval on each axis to explore interesting features. For example, Zhang and Kaufman~\cite{Zhao:2010:MRA} introduced a PCP to design high-dimensional transfer functions for identifying features by specifying the range between several scalar values. Liu and Shen~\cite{Liu:2016:AAF} drew PCPs independently for several interesting variables to visualize numerical distributions, and the axes of corresponding variable present its associated attributes. Then, the associated attributes of each variable were constrained to filter trivial data, and a high-dimensional transfer function was designed for extracting features. There are also a lot of studies that provide a basis for the application and improvement of PCPs~\cite{Guo:2012:SMV,Liu:2016:AAF,Zhou:2017:IPC} to explore the feature distributions of multivariate data. A SP is another means of representing the relationships among variables by dimension reduction methods, such as multidimensional scaling (MDS)~\cite{Kruskal:1978:MS} and t-SNE~\cite{Maaten:2008:VDE}, to identify and select a cluster by projecting data points onto a 2D visual plane. For instance, Guo et al.~\cite{Guo:2012:SMV} presented a scalable system that combines PCP and MDS projection to effectively identify features among variables, which shows the advantages of numerical distributions and cluster distributions. The abovementioned methods only take numerical distributions into account, but multivariate data usually has multiple features with the same numerical distribution but located in different regions, that is, spatial distributions, which are also significant for multivariate data.

A scatter plot matrix is another common interactive classification tool for guiding users to classify and extract features by projecting all variable pairs into a plot in a matrix form. For example, Guo et al.~\cite{Guo:2014:SLA} presented a visualization method for the analysis of multivariate unsteady flow data by the Lagrangian-based attribute space project (LASP) and provided a scatter plot matrix to project all the variable pairs of the selected pathlines. Lu and Shen~\cite{Lu:2017:MVD} proposed a bottom-up approach that employ a scatter plot matrix to project all variable pairs onto a 2D plane to identify features, as shown in Fig.~\ref{fig:sec2_1new}(b). In this work, users can define and modify features by selecting different 2D transfer functions iteratively.

This type of feature classification allows the flexible specification of features, and users are free to choose variables to define interesting features. Moreover, the quality of the extracted features could be higher with the participation of an experienced user. Certain features in multivariate data are often associated with only certain variables, and thus it is not necessary to analyze all the variables; this leads to a reduction in the computational overhead. In addition, finding all the meaningful features interactively remains a time-consuming and challenging task, and a lack of guidance for those users without prior knowledge of corresponding data sets makes it difficult to explore and classify the features.

\subsection{Feature Classification based on Data Mining}

Automatic or semi-automatic feature classification can be implemented by means of data mining methods~\cite{Han:2011:DMC} such as dimensionality reduction, clustering, and feature point matching to improve the analysis efficiency of multivariate data. In general, it is more difficult to design high-dimensional transfer functions for multivariate data, because the design often relies on user interaction with prior knowledge. In order to achieve good results, users need to perform ``trial-and-error'' work according to the statistical characteristics of the data, making the design process slow and inefficient. The feature classification based on data mining can avoid excessive interactions and improve the efficiency of exploration, which is classified into two categories as follows. The first is the visualization of interesting features selected from all features generated by automatically clustering. The second is the detection of similar features from the feature expression of a sample defined by users.

Global clustering can provide complete features in cluster space, from which users can choose interesting features to explore the characteristics of the data. Clustering is a common classification method that clusters the spatial data automatically or semi-automatically according to the correlation among voxels to identify different features in a 2D spatial plane. For scalar fields, Tzeng et al.~\cite{Tzeng:2004:ACV} employed the Iterative Self-Organizing Data Analysis Technique (ISODATA) clustering algorithm to generate classification results interactively. For vector fields, Hong et al.~\cite{Hong:2014:FLDA} presented a feature classification approach based on the Latent Dirichlet allocation (LDA) model by clustering pathlines with probabilistic assignment.
For multivariate data,
Wu et al.~\cite{Wu:2015:EAF} proposed an automatic analysis system integrating clustering to explore the multivariate data, visual encoding, and an interactive interface. In this system, different features can be classified automatically through a clustering-projection-classification iteration.

For the detection of similar features, subsets of distributions or regions of raw data are usually collected as initial features, from which similar features can be detected by means of multiple methods. At present, there are many mature feature classification methods, such as Gaussian naive Bayes, k nearest neighbor, support vector machines, neural networks, and random forests~\cite{Soundararajan:2015:LPT}. All these methods collect parts of the spatial distributions of raw data as initial classified samples, then learn different categories of transfer functions using machine learning algorithms, and finally apply these categories to the whole volume data. In this way, they avoid the complex interactive design of transfer functions and classify the features of raw data semi-automatically or automatically. Zhou and Hansen~\cite{Zhou:2014:GSS} proposed a semi-automatic exploration method for multivariate data that collects the regions of interest from different slices and extracts the features of corresponding regions automatically, as shown in Fig.~\ref{fig:sec2_1new}(c). In this work, the spatial distributions of multiple variables can also be fused for visualization. Wei et al.~\cite{Wei:2017:EDF} introduced two high-performance and memory-efficient algorithms for searching features that are characterized by marginal and joint distributions in multivariate fields. These algorithms leverage bitmap indexing and local voting to extract features that match a target distribution by the origin results and redefinition to generate all the matching results. Wang et al.~\cite{Wang:2016:MPM} presented a pattern matching method for multivariate data by the 3D scale-invariant feature transform (SIFT) algorithm with full rotation invariance to extract sparse sets of features for multiple scalar fields, from which a region of interest can be defined by users as SIFT features. The matching patterns in the entire data set are then located and ranked automatically.

Feature classification based on data mining is straightforward and effective, avoids excessive interactions, and makes it convenient to generate features in cluster space. However, its performance relies heavily on the clustering results or the feature classification algorithms, and it restraints user participation, which may contribute to the generation of unpredictable features.
In recent years, the relationship between machine learning and scientific visualization has become increasingly close. However, there are still vacancies for related work involving multivariate data. The design of transfer functions and feature classification are always difficult problems for multivariate data. Therefore, the identification of methods to utilize emerging technologies to solve the abovementioned problems of this field is one of the important avenues for future research.

\subsection{Feature Classification based on Topological Structures}
\label{Section_TopologicalStructures}

Surface features are also salient in spatial data fields. Feature classification based on topological structures is similar to the isosurface extraction of univariate spatial data. It uses abstract topological representation methods to extract surface features between multiple variables.

In multivariate data, the isosurface based on Marching Cubes in univariate spatial data can be extended to fiber surfaces (the isolines of two fields). Nagaraj and Natarajan~\cite{Nagaraj:2011:RIE} proposed a variation density function that profiles the relationship between multivariate fields over the isosurfaces of a given scalar field to guide users to select an isosurface of a variable that can strongly represent the change in other variables. Carr et al.~\cite{Carr:2015:FSG} extracted fiber surfaces in bivariate fields based on Marching Cubes to generate well-defined geometric surfaces, and analyzed the captured geometrical characteristics quantitatively, as shown in Fig.~\ref{fig:sec2_1new}(d). This feature classification method is similar to the high-dimensional transfer functions of volume rendering for multivariate data, although it fails to analyze the variable sets with more than two variables. Tierny and Carr~\cite{Tierny:2017:JFS} provided a practical and efficient algorithm extended from the concept of the univariate case for computing and extracting the Reeb space of bivariate data. The algorithm identifies the Jacobi set of bivariate fields (similar to the critical points in univariate spatial data), and then uses Jacobi Fiber Surfaces (similar to critical contours in univariate spatial data) to identify and compute the bivariate Reeb spaces. To improve the rendering efficiency,  Wu et al.~\cite{Wu:2017:DMV} presented an efficient direct ray casting method to render the fiber surfaces of bivariate spatial data in real time involving the computation of a distance field or the definition of control polygons.

A contour tree, an abstract topology representation for scalar fields, can be used to capture the variation of level sets in scalar fields~\cite{Boyell:1963:HTF}, which is also a common method to identify salient isosurfaces. Carr and Duke~\cite{Carr:2014:JCN} introduced a Joint Contour Net, a range-based quantization approach, to extract topological structures by quantifying the variation of multiple variables for topological analysis and visualization in multivariate domains, which has been applied to nuclear physical~\cite{Duke:2012:VNS} and meteorological simulation~\cite{Geng:2014:VAO}. They also reported an algorithm to construct joint contour nets for multivariate fields and introduced its theoretical and practical properties in detail.

Feature classification based on topological structures can efficiently extract the surface features between variables. However, unlike the topological structures of univariate data, it is difficult to clearly define the critical point for the topological structures and topological features of multivariate data. Therefore, methods to define and simplify the topological structures of multivariate data, and clearly identify the surface structures between variables are among the currently important research problems. Moreover, although there are many studies on the feature extraction of surfaces, they can only extract features from bivariate data at present, and thus the construction and identification of topological structures among three or more variables is still an urgent problem to be solved.

\section{Fusion Visualization}
\label{Section_Fusion}

\begin{figure*}[htbp]
\centering
\includegraphics[width=0.95\textwidth]{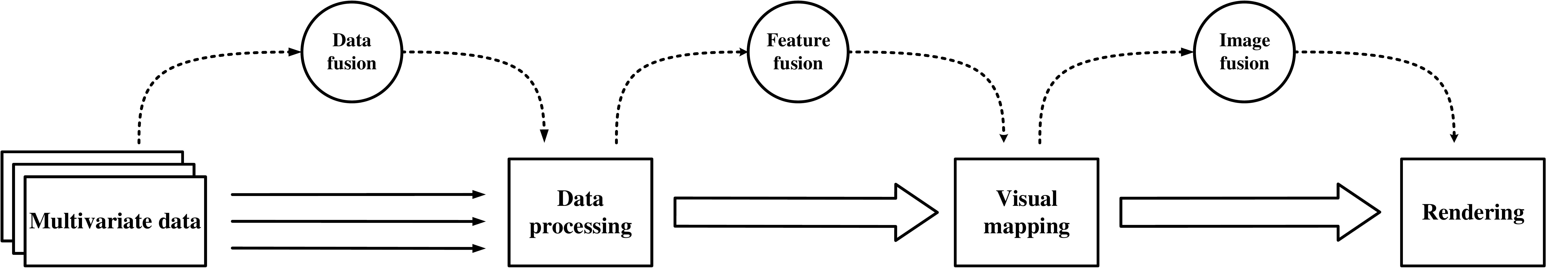}
\caption{Several fusion rendering processes of multivariate data. Data fusion is designed to extract features directly before data processing, whereas feature fusion takes place after data processing. After rendering and processing each variable separately, image fusion can be executed to fuse all the images.}
\label{fig:sec3_0}
\end{figure*}

The visualization of multivariate data was first proposed by Carwfis~\cite{Crawfis:1996:MVR}. Fuchs and Hauser et al.~\cite{Fuchs:2009:VOM} summarized the relevant research related to the visualization of multivariate scientific data before 2009. Features from different variables require simultaneous visualization to establish the spatial relationships among multiple variables. Conventional visualization methods for multivariate data can be classified into two categories: juxtaposed views and fusion visualization. For the former, Liu and Shen~\cite{Liu:2016:AAF} controlled the correlated thresholds of a variable and selected an interesting numerical distribution, and then the spatial distributions of other variables are presented side by side under the same thresholds. For the latter, Rocha et al.~\cite{Rocha:2017:DRL} proposed a real-time technique to map decals on surfaces to presenting multivariate data. In general, juxtaposed views fail to intuitively establish the corresponding relationships of several variables for same spatial points, so the range of application is relatively narrow. Fusion visualization is a widely used rendering method that maps different variable domains to different visual channels, such as colors, textures, opacity and icons, and then fuses these channels in a reasonable way.

For fusion visualization, there are two main challenges. One of these is \textbf{variable-related}. Multiple colors for different features are presented at each 3D spatial sampling point; when a new color is produced visually, the original colors may disappear or become too weak to presented. The other is \textbf{depth-related}. The rays are sampled for color blending and opaque blending, which can result in visual confusion and misleading results. For example, RGB channels can be used to represent three variables, but this may mislead users because the color blending of RGB may produce ambiguous colors.

In this section, we classify fusion visualization into three categories based on the visualization pipeline: data fusion, feature fusion, and image fusion, as shown in Fig.~\ref{fig:sec3_0}. Furthermore, we discuss the advantages and disadvantages of the three categories.

\begin{figure*}[hbp]
\centering
\includegraphics[width=0.95\columnwidth]{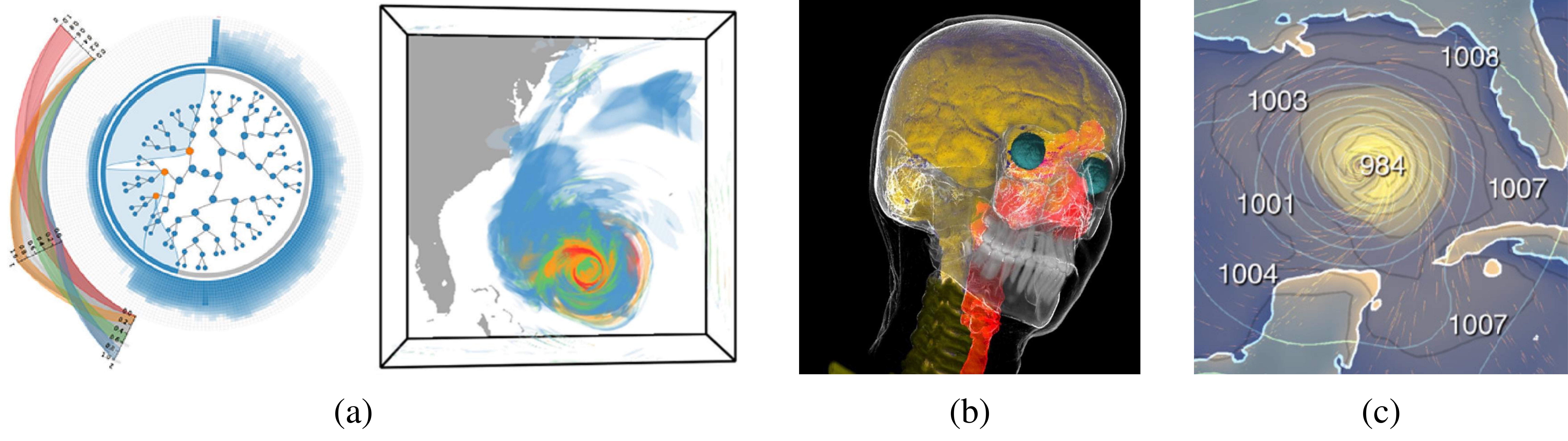}
\caption{The examples for fusion visualization. (a) The visual exploration of uncertain data sets using data fusion~\cite{He:2017:RLT}. (b) A two-level volume rendering method of segmented data sets using feature fusion~\cite{Hadwiger:2003:HTV}. (c) The interactive interface for creating multivariate time-varying data visualizations using image fusion~\cite{Schroeder:2016:VAA}.}
\label{fig:sec3_1}
\end{figure*}

\subsection{Data Fusion}
\label{Section_DataFusion}

For data fusion, the attributes of multivariate data can be processed directly, and certain data processing methods, such as statistical analysis and feature extraction, can be organically combined for visual analysis. Data fusion is represented by:
$
f(v_{1}(p), v_{2}(p), ..., v_{n}(p))\mapsto c(p),
$
where $v_{i}(p)$ and $c(p)$ are the scalar values of $i$-$th$ variable and the color of the sampled point $p$ respectively, $n$ is the number of variables, and $f$ is a mapping function to generate corresponding colors to distinguish different features. In general, we use $f$ to construct a new field from multiple variables or a high-dimensional transfer function for fusing multiple variables. Because the source, scale, or type of multivariate data may be different, data fusion needs to handle fusing multiple variables. In addition, multiple variables can also be processed directly for extracting salient features by statistical characteristics or distribution characteristics. Owing to the existence of various feature classification methods, there are various ways of fusing data.

A new field could be generated by fusing multiple variables to explore the relationships among variables. In this field, each spatial point represents the similarity or correlation of the variables. For instance, Dutta et al.~\cite{Dutta:2017:PIG} created a new scalar field, i.e., the pointwise mutual information (PMI) field, to preserve the combined and complementary information of multiple variables. They also extended the PMI field to a time-varying PMI field to capture the information from multiple time steps. In addition, topological structures in multivariate data, such as fiber surfaces, can usually be extracted by fusing multiple fields to improve the understanding of multivariate data. For example, Carr et al.~\cite{Carr:2015:FSG} extracted fiber surfaces from bivariate data. However, with the increase in the number of variables, the occlusion between variables and confused visualization make it difficult to explore these features. For this reason, Huettenberger et al.~\cite{Huettenberger:2014:DAS} proposed two techniques for simplifying the corresponding structures generated by Pareto sets for multivariate data. The first technique is based on a weighted graph to present connected components of Pareto extrema. The second simplification technique follows the lines of Suthambhara and Natarajan~\cite{Suthambhara:2011:SOJ} to compute and simplify the contour tree for multivariate data. In particular, the analysis of more than two variables is supported in this work.

Multiple variables can also be fused by high-dimensional transfer functions without considering spatial information. In general, high-dimensional transfer functions are designed interactively. For example, Lu and Shen~\cite{Lu:2017:MVD} designed a bottom-up method for interactive subspace exploration to extract inner features of multivariate data, and then mapped these features to different colors, as shown in Fig.~\ref{fig:sec2_1new}(b). Guo et al.~\cite{Guo:2012:SMV} designed high-dimensional transfer functions combining a PCP and multi-scaling projection, and used color coding of multiple features to fuse multiple variables before data processing. He et al.~\cite{He:2017:RLT} provided an improved PCP for designing high-dimensional transfer functions based on Range Likelihood Tree to code colors to render different features, as shown in Fig.~\ref{fig:sec3_1}(a).

There is a noise-based volume rendering method for multivariate data in which only one variable is calculated for each pixel in a 2D image by a noise-based ray casting algorithm. The method can avoid the two problems mentioned above by coding different color channels without color blending, which is also a common method of data fusion. The noise-based method was first proposed by Carwfis~\cite{Crawfis:1996:MVR}. They regarded volume density cloud and noise amount as the first and second variables, and proposed noise splats for fusing the relationships between variables, making different variables clearly distinguishable with high resolution. Khlebnikov et al.~\cite{Khlebnikov:2013:NVR} proposed a method for multivariate data visualization that generates an opacity redistribution pattern and an opacity mapping function with high-frequency to avoid confusion by random-phase Gabor noise. After this, different variables can be rendered by different colors, which allows users to distinguish different distributions with high resolution.

The noise-based volume rendering method can avoid the occlusion among variables and confused visualization for multivariate data in a single view, which is a better visualization method than others. However, the method relies on many structure models, the information carrying capacity is limited, the depth information is missing, and the visualization results depend on the design of the noise functions.

\subsection{Feature Fusion}
\label{Sec_FeatureFusion}

For feature fusion, features from each variable are usually extracted and expressed, such as the isosurfaces of multiple variables, the fusion of isosurfaces or colors in scalar fields and streamlines in vector fields, and the different regions rendered by different variables. Feature fusion can be expressed as:
$
g(f(v_{1}(p))\mapsto c_{1}, f(v_{2}(p))\mapsto c_{2},  ..., f(v_{n}(p))\mapsto c_{n})\mapsto c(p),
$
where $f$ represents the corresponding mapping function for each variable to obtain its respective color $c_{i}$, $n$ is the number of variables, and $g$ represents the fusion operation to obtain the final color. Usually, $g$ can mix multiple colors and opacities to produce a new color, i.e., color blending. Moreover, $g$ can also control the presentation of each variable. For example, the features or variables are hidden when the opacity is set to 0. For this method, the visual depth information among features can be preserved.

Color blending is the most commonly used method for visualization~\cite{Akiba:2007:VMV,Akibay:2007:ATV}. Color blending can be further classified into the fusion based on the integrating process of volume rendering and result-based image fusion. The fusion based on the integrating process of volume rendering can present the spatial distributions of features intuitively by fusing the colors of different features at each sampling point, whereas new colors generated by the fusion may easily cause ambiguity and even mislead users into making incorrect judgements. For these problems, Ding et al.~\cite{Ding:2016:VIO} presented a method that employed multi-class sampling to individually generate and illustrate spatial sampling points for each variable. Three visualization modes were provided for spatial distributions and correlations of multiple variables, which can identify the colors and variables more accurately. The method is similar to color weaving~\cite{Urness:2003:EVM,Hagh:2007:WVB}. Color weaving generally displays multiple original independent colors in a high-frequency texture side by side to fill corresponding regions, which can deal with the overlapping of multiple colors and overcome the disadvantage of traditional color blending.

Different features can also be rendered by different colors to present the relationships among variables. The challenges of occlusion, clipping and fusion for conventional multivariate visualization make it difficult to visually explore the features of multiple variables. For this problem, Chuang et al.~\cite{Chuang:2009:HCB} and Kuhne et al.~\cite{Kuhne:2012:ADA} successively proposed hue-preserving color blending methods to avoid false colors and preserve the depth information of multiple features in multivariate scalar fields. In multivariate vector fields, Guo et al.~\cite{Guo:2014:SLA} proposed a scalable method for the exploration of multivariate unsteady flow data with LSAP. The method projects pathlines starting from spatial points onto a 2D plane to reduce the complexity and incorporates a MapReduce-like framework with scalable Pivot MDS to simplify the field line tracing, from which users can select groups of points and the corresponding pathlines can be presented.

The main boundary structure can be clearly presented through surface rendering, and volume rendering is expected to display the internal features of objects. The combination of the two rendering methods can make full use of their respective advantages, which can further depict the phenomena and laws hidden in data, and improve the ability to comprehensively display images. In many applications, the results of volume rendering and surface rendering need to be displayed simultaneously to achieve the optimal display of features. Hybrid rendering is a technique to combine the two. For example, in the seismic field, it is necessary to show the fault plane, the surface of reservoir rock, and other geologic formations in sequence to design the drilling trajectory and select the optimal mining plan. Hauser et al.~\cite{Hadwiger:2003:HTV} proposed a two-level volume rendering technology with which one of the specific rendering methods including maximum intensity projection, direct volume rendering, and isosurface rendering can be selected for each subregion to achieve feature fusion, as shown in Fig.~\ref{fig:sec3_1}(b). Kreeger and Kaufman~\cite{Kreeger:1999:MTP} realized feature fusion through sampling 3D textures and depth buffers. In this work, surface rendering was used to render blood vessels, and volume rendering was used to render other tissues for visualizing the MRI data of a patient's head.

Feature fusion can give full play to the fused features of different variables or those among variables, and it can avoid producing new colors and preserve the depth information among features, making it possible to visualize multiple features more intuitively. However, visual occlusion may still occur when there are too many fused features, and the quality of the visual results relies on structural patterns and feature classification methods.

\subsection{Image Fusion}

For image fusion, each variable is rendered to an image or visual primitive (hereinafter collectively referred to as image) through different visual channels, and then the images are fused for visualization. Each pixel of the resultant image only presents the color of one image. The expression of image fusion is
$
I(Image_1, Image_2, Image_3, ...)\mapsto Image,
$
where $Image_i$ represents the result of the $i$-$th$ variable, and $I$ is the fusion function of multiple results to get the final $Image$. Image fusion is simple and intuitive, and makes it easy to explore the inner relationships of multivariate data.

Section~\ref{Sec_FeatureFusion} discusses the two categories of color blending. For result-based image fusion, multiple images are fused directly to get the final image. For instance, Zhou and Hansen~\cite{Zhou:2014:GSS} provided lasso tools to select regions of interest as initial features from multiple slices, and similar features are extracted to show the 3D spatial distributions based on image fusion.

An icon-based method encodes different variables or features into different tuples to distinguish them through human perception of shape, size, direction, etc. It can transmit direction information well, and thus it is widely used in the visualization of vector fields~\cite{Kirby:1999:VMD} and tensor fields. However, icons are easily obscured, making it difficult to apply to multivariate data. Compared to the icon-based method, texture-based coding can provide a more natural and compact structure by mapping different colors or opacity to this structure to achieve the goal of fusing multiple variables at the same time. Urness et al.~\cite{Urness:2003:EVM} proposed the concept of color weaving in 2D multivariate flow fields. They assigned different colors to different streamlines in the texture generated by the Line Integral Convolution (LIC), such that multiple colors could be represented simultaneously at the same spatial point. Schroeder and Keefe~\cite{Schroeder:2016:VAA} provided a sketch for origin visualization, through which users can interactively and directly create new time-varying multivariate data visualization by means of rich tools to map colors or icons, as shown in Fig.~\ref{fig:sec3_1}(c). Glyph-based visualization~\cite{Borgo:2013:GVF}, similar to the icon-based method, usually uses glyphs such as shapes, indices, and textures to place the locations of variables, which is widely used in 2D, 2.5D, or 3D visualization space.

In conclusion, color blending is widely used for its intuitive perception, but ambiguous colors or mutual occlusion among features makes the information of variables difficult to distinguish. The icon-based and texture-based methods can avoid occlusion and visual confusion to some extent, though they require users to recover lost information according to a random structure, which increases the burden of user perception. Noise-based methods can be applied to avoid this problem. Moreover, focus and context visualization makes it possible to retain contextual information when local features are displayed in the focused area~\cite{Kruger:2006:CAI}, which provides an idea to solve the above problem in multivariate spatial visualization.

\section{Correlation Analysis}

In general, the relationships between variables of multivariate data are intricate and complex. According to the objectives of correlation analysis, we divide the correlation into five different categories: correlation between variables, correlation between voxels, correlation between numerical values, correlation between features and the hybrid analysis method for variables and numerical values.

\begin{figure*}[htbp]
\centering
\includegraphics[width=0.95\columnwidth]{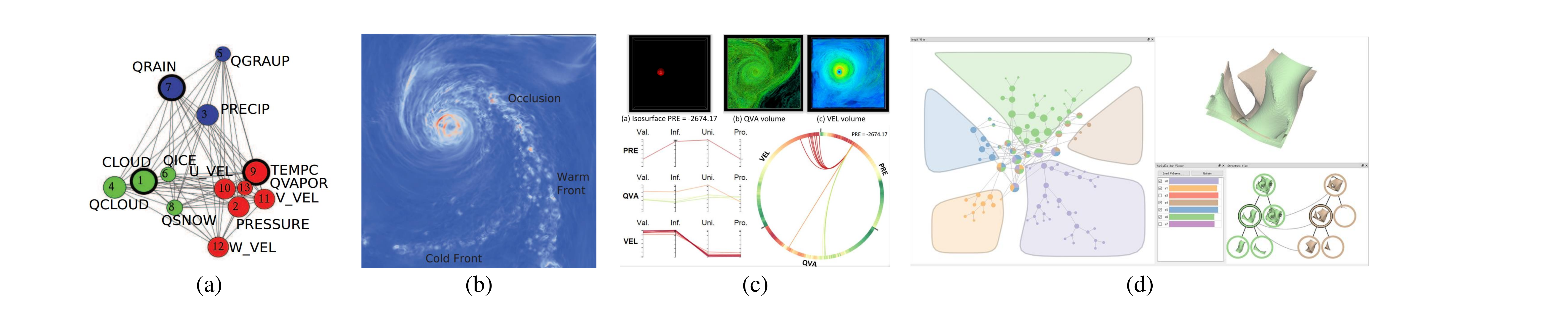}
\caption{Examples of multivariate correlation analysis. (a) Correlation between voxels measured by gradient~\cite{Nagaraj:2011:AGC}. (b) Correlation between variables measured by mutual information~\cite{Biswas:2013:AIF}. (c) Correlation between numerical values based on the associated relationships~\cite{Liu:2016:AAF}. (d) Correlation between features based on FeatureNet~\cite{Wang:2017:FAV}.}
\label{fig:sec4_1}
\end{figure*}

\subsection{Correlation between Voxels}

The correlation between voxels measures the similarities of scalar values or their derived attributes.

The correlation measure for each voxel is mostly based on the similarity of gradient fields. Gosink et al.~\cite{Gosink:2007:VII} measured the gradient correlation based on the cosine of two gradient vectors, i.e., a related scalar field of two gradient fields was generated by the inner product of each spatial sampling point. Sauber et al.~\cite{Sauber:2006:MAA} introduced a similarity measure method named gradient similarity measure (GSIM), as follows:
\begin{equation}
s(\triangledown f_{i}, \triangledown f_{j})=(s_{d}(\triangledown f_{i}, \triangledown f_{j})\cdot s_{m}(\triangledown f_{i}, \triangledown f_{j}))^{r},
\end{equation}
\begin{equation}
s_{d}(\triangledown f_{i}, \triangledown f_{j})=(\frac{\triangledown f_{i}^{T}\triangledown f_{j}}{\left \| \triangledown f_{i}\cdot \triangledown f_{j} \right \|})^{2},
\end{equation}
\begin{equation}
s_{m}(\triangledown f_{i}, \triangledown f_{j})=\frac{4\cdot \left \| \triangledown f_{i} \right \| \cdot \left \|\triangledown f_{j} \right \|}{(\left \| \triangledown f_{i}\ \right \| + \left \| \triangledown f_{j} \right \|)^{2}}.
\end{equation}
The above expressions take into account the similarity of magnitude and direction of two gradient fields based on voxels, where $s_{d}$ and $s_{m}$ represent the similarity in direction and magnitude respectively, $\triangledown f_{i}$ represents the gradient with respect to the $i$-$th$ variable, and the exponent $r$ regulates the sensitivity of the measure.

All the above methods aim at the measurement of bivariate fields. Sauber et al~\cite{Sauber:2006:MAA} then proposed a method to measure the gradients of all the variable pairs by the minimum similarity, extending GSIM to multiple variables. Nagaraj et al.\cite{Nagaraj:2011:AGC} introduced a gradient-based comparison method for multivariate data. The method is defined as a partial derivative matrix comprising the gradient vectors of different fields. Let $F=\{f_{1}, f_{2}, ..., f_{m}\}$ be a set of scalar fields on a manifold $\mathbb{M}$. The differential at point $p\in\mathbb{M}$ is defined as:
\begin{equation}
dF(p)=\begin{bmatrix}
\frac{\partial f_{1}}{\partial x_{1}}(p) & ... & \frac{\partial f_{1}}{\partial x_{n}}(p) \\
... & ...& ...\\
\frac{\partial f_{m}}{\partial x_{1}}(p) & ... & \frac{\partial f_{m}}{\partial x_{n}}(p)
\end{bmatrix}.
\end{equation}
The multi-field comparison measure $\eta _{p}^{F}$ is defined as $\left \| dF(p) \right \|$, which represents the norm of the matrix $dF(p)$ at sampling point $p$. This method calculates the correlation between variables at each voxel point to generate a derived scalar field, as shown in Fig.~\ref{fig:sec4_1}(a). Usually, the new scalar field needs to be classified to identify salient regions with strong relationships.

Clustering is also a common method to analyze the correlation between voxels. There are many methods that use cluster algorithms in the field of data mining to classify similar voxels into one group for identifying features~\cite{Tzeng:2004:ACV,Linsen:2009:MIV,Wu:2015:EAF}.
Van Long and Linsen~\cite{Linsen:2009:MIV} proposed a hierarchical density-based clustering method, in which the density is calculated based on grid cells. Compared with other clustering methods, the main advantage of this method is that the thresholds of the density do not need to be specified. The authors also designed a radical layout to intuitively show the results of hierarchical clustering. In the correlation analysis of voxel-based methods, visual analysis methods prefer to combine human intelligence to obtain more accurate clustering results. The simplest idea is to provide a rough clustering result of the data, and then iteratively correct the clustering result in an appropriate way. Ivanovska and Linsen~\cite{Ivanovska:2008:AUT} clustered multivariate data with certain simple and efficient methods such as k-means and median cut, and provided a 2D slice view to correct the clustering results by splitting and merging operations.

The similarity between voxels is defined based on all variables, such the correlation between variable might avoid missing important features. However these methods cannot identify the features only correlated with partial variables. Moreover, irrelevant variables may have a great influence on clustering results with the increase in the number of variables, which may easily mislead users to explore trivial features.

\subsection{Correlation Between Variables}

The overall correlation measurement of variables mainly considers the interaction between variables. Statistical analysis and information theory are introduced in this section.

The correlation coefficient is a standard and common statistical measurement method that determines whether two variable sets are linearly dependent by comparing the differences between their scalar values and their respective mean values. As a global measurement method, the correlation coefficient can also be used to calculate the local correlation between two variables~\cite{Sauber:2006:MAA}. For time-varying data, a linear relationship for the same spatial sampling point between two time steps can also be measured by the correlation coefficient~\cite{Sukharev:2009:CSO}.

Information theory provides a theoretical framework for the measurement of the amount of information between variables, which is widely used in the visualization field to measure the importance of a variable or the similarity between two variables. The correlation between two variables can be measured through the amount of information shared by the two variables, i.e., mutual information. Biswas et al.~\cite{Biswas:2013:AIF} employed mutual information to measure the informativeness of one variable about the other variable and grouped variables based on mutual information in a graph-based approach, as shown in Fig.~\ref{fig:sec4_1}(b). Because there is information redundancy between variables, they first clustered and grouped variables using mutual information as a measurement of distance, and then used conditional entropy to determine the variable with minimum uncertainty as the most important variable for each group to further analyze the correlation between variables. In addition, information theory can also be extended to time-varying fields. For example, Wang et al.~\cite{Wang:2011:AIT} applied the transfer entropy to investigate the salient relationships between variables in time-varying multivariate fields. Dutta et al.~\cite{Dutta:2017:PIG} extracted important features by mutual information and its two decompositions in time-varying multivariate data, and multiple time-varying features were encoded into the same individual field to analyze the track and characteristics of these features.

The correlation between variables is simple and straightforward, and it usually measures the global correlation between two variables. However, as most of the correlation between variables is based on the overall correlation between two or more variables, it is difficult to extract the significant features of local correlation between variables. Moreover, the correlation between variables fails to take the composition of the numerical and spatial distributions of variables into account.

\subsection{Correlation Between Numerical Values}

The correlation between numerical values is used to analyze the relationships of value pairs for multiple variables.

A PCP can map multiple variables onto the same 2D plane to present the numerical relationships intuitively, which is a very common numerical analysis method in the study of multivariate data. However, owing to layout constraints, only the relationship between adjacent axes can be intuitively represented, and mutual occlusions between coordinate axes make it difficult to identify correlation patterns between numerical values for large-scale data. In addition, it is impossible to interactively select all possible correlation patterns in a PCP, and thus many automatic numerical analysis methods are proposed. Haidacher et al.~\cite{Haidacher:2011:VAU} established distance fields for the isosurfaces of variable domains, and measured these distance fields using mutual information to calculate the similarity between isosurfaces. A similarity matrix was generated to show all the equivalent pairs of two variables in this work. Liu and Shen~\cite{Liu:2016:AAF} extracted the relationships between scalar values in two different variable domains and used the Influence-Passivity Model to choose the most representative scalar value, as shown in Fig.~\ref{fig:sec4_1}(c).

Topological structures can also be applied to the correlation analysis of numerical values. Contour tree~\cite{Carr:2003:CCT} is used to represent the relationship between values according to the change in level sets. Carr and Duke~\cite{Carr:2014:JCN} extended the contour tree of multivariate data to establish topological relationships between different numerical intervals in high-dimensional data. In addition, various topological structures for multivariate data are proposed to measure the changes between values such as the Jacobi set~\cite{Edelsbrunner:2002:JSO}, and Pareto set~\cite{Huettenberger:2013:TMS}.

The correlation between numerical values can be used to analyze the global and local relationships by flexibly controlling the parameters of each variable. In this method, the numerical values of each variable are employed to analyze the numerical distributions. However, it fails to take the spatial distributions into account, and thus some trivial features with spatial distributions may be ignored.

\subsection{Correlation between Features}

The correlation between features aims at analyzing the differences and similarities between the features of multiple variables.

Schneider et al.~\cite{Schneider:2008:ICO,Schneider:2013:ICO} presented an efficient method that compares different features between variables based on simplified contour trees. They extracted multiple largest contours of each variable domain, calculated the similarity between the contours, and finally extracted the similar contours of different scalar fields by clustering. The method effectively avoids the ambiguous features defined by transfer functions in the process of rendering multiple surface meshes. Wang et al.~\cite{Wang:2017:FAV} proposed an interactive visual analysis method for the features and relationships in multivariate data. They extracted the main features between variables and the hierarchical relationships between these features based on merge trees. Next, they established the associated relationships between the features by defining the similarities between them to merge the feature trees of different variables. A FeatureNet was then constructed as a navigation tool to guide users to explore the correlation between features, as shown in Fig.~\ref{fig:sec4_1}(d).

The correlation between features takes the local relationships between features into account, and it is easy to clearly identify the different features. However, the correlation results rely on the feature classification methods.

\subsection{Value-variable Correlation Analysis}

The several correlation analysis methods mentioned above, whether based on the global correlation analysis for variables and voxels or based on the local correlation analysis for numerical values and features, only analyze one aspect for multivariate data. However, variable sets usually show strong correlation in multiple aspects. To address this problem, the value-variable correlation analysis can avoid the one-sidedness of the above methods. For example, Biswas et al.~\cite{Biswas:2013:AIF} employed mutual information to measure the informativeness of one variable about another variable. In this work, a hybrid correlation analysis combining numerical values and variables was employed for exploring features in multivariate data, because the corresponding isosurfaces were selected for the measurement of scalar values and other variables were also be presented for the measurement of variables. Liu and Shen~\cite{Liu:2016:AAF} analyzed the correlation between numerical values using PCPs, and then selected an interesting isosurface of a variable. The volume rendering results for other variables were presented under the condition of side-by-side isosurfaces. In this work, numerical values and variables are analyzed, which avoids the complexity of subspace exploration. In addition, features from multiple variables can be independently filtered according to the feature extraction methods of univariate data, and then feature fusion is employed to analyze the correlation between features and variables. The value-variable correlation analysis method avoids the one-sidedness of unilateral correlation, and is strongly related to the feature classification of multivariate data.

\section{Conclusion}

Multivariate data visualization is designed to efficiently express and analyze variables and their potential relationships, and then display and explore the evolution laws of complex scientific phenomena by visualization to assist scientists in discovering new relationships or laws. Therefore, multivariate data visualization has always been an important research topic in the field of scientific visualization. In this paper, we provide a comprehensive survey of multivariate data visualization with a focus on features classification, fusion visualization, and correlation analysis. Many advanced works have been developed with the respect to the above three aspects in different fields, such as medicine, electromagnetism, combustion simulation, and meteorological simulation. With the rise of applications in large-scale scientific and engineering fields, the devices and methods of data acquisition have become increasingly intelligent and wide-ranging, bringing great challenges and opportunities to multivariate data visualization. Certain potential research topics for multivariate data visualization according to current research are discussed below.

For \textbf{feature classification}, it is essential to locate, extract, and analyze important features of multivariate data, whereas it is time consuming and challenging to find all the meaningful features interactively if the user has little prior knowledge. Deep learning has been a hot topic in recent years, although it has few applications in multivariate data analysis. It can be used to automatically train and express multivariate data and implement the classification by intelligently combining spatial and numerical distributions, which are considered potential future research topics. Moreover, feature classification based on topological structures can only extract the features of bivariate data at present, and thus the construction and identification of topological structures among three or more variables are still urgent problems to be solved. In the time-varying multivariate field, a time-varying feature may exist throughout time steps, so the feature classification methods in multivariate data aiming to extracting multiple features of one time step can also be extended to extract and trace time-varying features in the future.


For \textbf{fusion visualization}, the rendering of univariate spatial data is relatively mature, whereas the fusion visualization for multivariate data still has problems such as visual confusion, occlusion, and fatal colors. There are many excellent methods to solve these types of problems, but these methods still have various defects within partial perspectives. Thus, the search for methods to solve the above problems efficiently and thoroughly is a promising direction for future research. Moreover, there are lots of methods for fusion visualization, whereas the quality of the perception-awareness of these methods fails to be considered. This brings an opportunity to explore the evaluation methods of fusion visualization.

For \textbf{correlation analysis}, the correlation between variables is focused on global analysis at present, whereas multiple data always have local relationships between partial variables. The analysis of the local relationships between variables is important for future exploration. In addition, the correlation between multiple time steps in time-varying multivariate data needs to be examined. Furthermore, multivariate data usually has a combined mode, that is, different objects only show strong correlation under a specific combination of variables, and certain variables are only related to portion of the objects. The correlation analysis in information visualization is efficient and rich to analyze the combine mode. For example, subspace analysis in information visualization can analyze the local relationships between attributes and data items, which could be applied to multivariate scientific data. For instance, He et al.~\cite{He:2018:BBV} presented a bicluster method to cluster variables and voxels simultaneously to extract all biclusters with a similar scalar value pattern automatically, which applies subspace analysis of information visualization to scientific visualization.

In recent years, the work of fusion visualization for multivariate data has gradually matured, and most of the work has focused on feature classification and correlation analysis. Due to the improvement in the computational performance of supercomputers, the number of variables and data size are becoming increasingly complex, and the complexity of multivariate data increases the difficulty of the understanding of the intricate relationships among multiple variables, i.e. feature classification and correlation analysis. This bring great opportunities in data representation to reduce the data size and analyze on compact data representation~\cite{Wei:2015:ELH,Wei:2018:IGD}. Moreover, finding methods to reconstruct the multivariate data reasonably, better classify features, and render the spatial fields of different variables remain problems that need to be addressed.


\section*{Acknowledgement}

The authors would like to thank the anonymous reviewers for their valuable comments. This work was supported by the National Key Research \& Development Program of China (2017YFB0202203), National Natural Science Foundation of China (61472354 and 61672452), NSFC-Guangdong Joint Fund (U1611263), and the Fundamental Research Funds for the Central Universities.

\bibliographystyle{spmpsci}      
\bibliography{template}   

\end{document}